\documentclass[11pt]{article}
    \usepackage{style}

    \usepgfplotslibrary{external} 
    \usetikzlibrary{positioning,fit,calc,decorations.pathreplacing,calligraphy, arrows.meta, pgfplots.groupplots}
\tikzstyle{Red firm}=[fill={rgb,255: red,216; green,27; blue,96}, draw=black, shape=circle]
\tikzstyle{Yellow firm}=[fill={rgb,255: red,255; green,193; blue,7}, draw=black, shape=circle]
\tikzstyle{Blue firm}=[fill={rgb,255: red,30; green,136; blue,229}, draw=black, shape=circle]
\tikzstyle{Hidden blue firm}=[fill={rgb,255: red,30; green,136; blue,229}, draw=black, shape=circle, dashed, opacity=0.5]
\tikzstyle{Not functional}=[fill=white, draw=black, shape=rectangle]
\tikzstyle{Green firm}=[fill={rgb,255: red,0; green,128; blue,32}, draw=black, shape=circle]

\tikzstyle{Dark blue firm}=[fill={rgb,255: red,0; green,32; blue,76}, shape=circle]
\tikzstyle{Almost dark blue firm}=[fill={rgb,255: red,99; green,102; blue,111}, shape=circle]
\tikzstyle{Black firm}=[fill={rgb,255: red,48; green,48; blue,48}, shape=circle]
\tikzstyle{Almost dark yellow firm}=[fill={rgb,255: red,239; green,216; blue,82}, shape=circle]
\tikzstyle{Dark yellow firm}=[fill={rgb,255: red,255; green,233; blue,69}, shape=circle]

\tikzstyle{Unseen edge}=[dotted, ->, opacity=0.5]
\tikzstyle{Potential supplies}=[dotted, -, line width = 1]
\tikzstyle{Supplies}=[-, line width = 1]

\tikzstyle{probability}=[fill={rgb,255: red,154; green,212; blue,214}, draw=black, shape=circle, inner sep=5, minimum size=35]
\tikzstyle{distribution}=[fill={rgb,255: red,255; green,186; blue,73}, draw=black, shape=circle, inner sep=1, minimum size=50]
\tikzstyle{empty}=[fill=white, draw=black, shape=circle, inner sep=1, minimum size=35]
\tikzstyle{dashed}=[fill=white, draw=gray, shape=circle, inner sep=1, minimum size=35]

\tikzstyle{implies}=[->, ultra thick]
\tikzstyle{sees}=[->, dashed, ultra thick]
\tikzstyle{follows}=[-, dotted, ultra thick]

    \pgfplotsset{compat=1.18}

    \newcommand{\inputpgf}[1]{%  
        \IfFileExists{#1}{\input{#1}}{
            \includegraphics[width=0.5\textwidth]{example-image-a}
        }
    }

    \usepackage{subfiles} % Load last

    \newcommand{\D}{\mathcal{D}}

    \renewcommand{\S}{\mathcal{S}}

    \renewcommand{\P}{\mathbb{P}}
    \newcommand{\E}{\mathbb{E}}
    \newcommand{\V}{\mathbb{V}\text{ar}}

    \newcommand{\abs}[1]{\left\lvert#1\right\rvert}

    \newcommand{\Beta}{\text{Beta}}
    \newcommand{\Bin}{\text{Bin}}

    \newcommand{\BP}{\text{BetaPower}}

    \newcommand{\ceil}[1]{\left\lceil#1\right\rceil}

    \author{Andrea Titton \\
    \href{mailto:a.titton@uva.nl}{\texttt{a.titton@uva.nl}}}
    \title{Risk Propagation in Endogenous Supply Chains \thanks{
      Faculty of Economics and Business, University of Amsterdam (a.titton@uva.nl). I thank my supervisors Florian Wagener and Cees Diks for the patient guidance on this paper. I also thank the CeNDEF group and the Quantitative Economics section at the University of Amsterdam, as well as the CES group of Paris 1, for helpful comments throughout the many seminars. Finally, I thank attendees of the EEA-ESESM conference, Barcelona, 2023, the Dutch Network Science Society Symposium, Leiden, 2022,for the constructive comments.
    }}
    \date{\today}

\begin{document}

\maketitle
\begin{abstract}
  This paper investigates the endogenous formation of supply chains and its consequences for disruption propagations. In production networks where upstream risk is highly correlated and supplier relationships are not observable, the marginal risk reduction of adding an additional supplier is low, because this additional supplier's risk is likely to be correlated to that of the firm's existing suppliers. This channel reduces firm incentives to diversify, which gives rise to inefficiently fragile production networks.

  By solving the social planner problem, I show that, if the risk reduction experienced downstream resulting from upstream diversification were to be internalised by upstream firms, endogenous production networks would be resilient to most levels of risk. Furthermore, I show that imperfect information yield inefficient but more robust supply chains. Despite its stylised form, the model identifies the trade-off firms face when diversifying risk and isolates the mechanism that aggregates these decisions into a production network. Furthermore, it maps the conditions of the trade-off, such as expected profits of the firm or the sourcing costs, to the properties of the production network. 
\end{abstract}

\newpage
% Introduction

In August 2020, hurricane Laura hit one of the world's largest petrochemical districts, in the U.S. states of Louisiana and Texas. As polymer producers in the area were forced to halt production, up to 15\% of the country's polypropene (PE) and polypropene (PP) producers were unable to source polymer inputs, which in turn caused shortages across the economy \citep{vakil_latest_2021}. Such a widespread disruption raised awareness on the role the correlation of suppliers' risk has in destabilising production networks and its importance in firms' sourcing decisions. In face of such correlated risk, how do producers make sourcing decisions? And, should we expect these sourcing decisions to yield resilient production networks or to amplify the idiosyncratic risk?

In this paper, I study the feedback between the risk of a disruption in sourcing inputs and the endogenous formation of supply chains. A widespread approach to mitigate risk is to diversify it by multisourcing. This practice consists of procuring the same inputs from multiple suppliers, sometimes redundantly \citep{zhao_robust_2019}. Yet, when deciding how many suppliers from which to source, a firm faces decreasing marginal benefits in risk reduction, because each additional supplier's failure to deliver is increasingly likely to be correlated with that of the firm's current suppliers. In the presence of marginal costs of sourcing, for example contractual costs or higher prices, the uncertainty behind the correlation of a firm's potential suppliers might induce it to diversify risk less than socially optimal. The wedge between endogenous firm decisions and social optimality arises because downstream firms would be willing to compensate their suppliers for increased diversification of inputs. This underdiversification can generate aggregate fragility in production networks. To understand the relationship between the firm's diversification decisions and supply chain fragility, I study the properties of a stylised production game. In the equilibrium of the game, correlation in the risk of disruption among suppliers generates fragility via two channels. First, it directly introduces endogenous correlation in downstream firms' risk, which amplifies through the production network. This increases the probability of cascading failures, in which the entire production network is unable to produce. Second, it indirectly affects firms' decisions by reducing the expected marginal gain from adding a source of input goods. The latter channel leads to firms diversifying increasingly less, such that small increases in the expected disruption probability can yield fragile production networks.

The role that production networks play in determining economic outcomes has been long recognised. As far back as \cite{leontief_quantitative_1936}, economists have studied how networks in production can act as aggregators of firm level activity. Following a foundational paper by \cite{hulten_growth_1978}, which showed that the first order impact of a productivity shock to an industry is independent of the production network structure, macroeconomics has since de-emphasised this role \citep[p. 2]{baqaee_macroeconomic_2019}. However, more recently, \cite{baqaee_macroeconomic_2019} illustrated how the structure of the production network can aggregate micro shocks via second order effects.\footnote{These results build on a vast literature and recent literature (e.g. \cite{gabaix_granular_2011, acemoglu_network_2012,carvalho_supply_2020, baqaee_macroeconomic_2019,carvalho_production_2019})} Furthermore, the degree of competition in an industry also interacts with the production network to aggregate shocks, which can lead to cascading failures \citep{baqaee_cascading_2018}. Once established that production networks play a central role in aggregating shocks, two natural questions arise. First, which networks can we expect to observe, given that firm endogenously and strategically choose suppliers? Second, are these endogenous network formations responsible for the growth or fragility that large economies display? These questions fuelled a number of recent papers studying endogenous production network formation. Focusing on growth, \cite{acemoglu_endogenous_2020} show that endogenous production networks can be a channel through which firms' increased productivity lowers costs throughout the supply chain and allows for sustained economic growth. In parallel, a vast literature dealt with studying the role of endogenous production networks and firm incentives in determining fragile or resilient economies. \cite{erol_network_2014} showed that in networks with strategic link formation, systemic endogenous fragility arises if the shocks experienced by firms are correlated. Later work, by \cite{amelkin_strategic_2020}, shows that uncertainty in the time of production is crucial in determining whether production networks in equilibrium are sparse, hence fragile. Finally, \cite{elliott_supply_2022} illustrate how complexity in the production process can also be a key driver of endogenous fragility in production networks. \footnote{The literature on production networks is vast and it is unfortunately impossible to give a fair overview in this introduction. For a more comprehensive review of the literature I refer the reader to \cite{carvalho_production_2019} and \cite{amelkin_strategic_2020}}

A less understood link is that between the opacity of the supply chain, how firms deal with it, and which consequences this has on the economy. \cite{kopytov_endogenous_2021} studied the effect of uncertainty in endogenous production network formation on firms' productivity and business cycles. They find that higher uncertainty can lead to lower economic growth. In contrast, this paper focuses on the role of uncertainty in generating endogenous fragility to cascading failures using a more stylised production network model, akin to that studied by \cite{elliott_supply_2022}. In line with the existing literature, in the model small idiosyncratic shocks can be massively amplified. The degree of amplification depends on the equilibrium behaviour of firms. This phenomenon holds true in vertical economies producing simple goods. The novel theoretical contribution of this paper is to extend the analysis of production network formation to an environment in which firms aim to minimise risk while accounting for correlation between suppliers. To do so, I develop a tractable analytical framework that describes the propagation of idiosyncratic shocks through the supply chain when firms take sourcing decisions endogenously in an imperfect information environment. The model describes the evolution of risk through the supply chain as a dynamical system over its depth. The social planner solution shows that endogenous fragility can impose large welfare losses. Importantly, these losses might be discontinuous: an arbitrary small increase in the correlation of risk among basal firms, can generate large welfare losses. Finally, I study a benchmark case where firms have perfect information over idiosyncratic risk. In this case, despite each individual firm being able to achieve smaller disruption risk, the supply chain is maximally fragile and there is a high probability of large disruptions. 

The remainder of the paper is structured as follows. Section \ref{section:structure} discusses the assumptions on the supply chain disruptions, the problem of the firm, and establishes the results that allow the firm to make sourcing decisions. Section \ref{section:propagation} derives the law of propagation of the disruption events through the supply chain. Section \ref{section:optimum} establishes the firm's optimal sourcing strategy and how this endogenously determines the fragility of the supply chain. These results are then compared, in Section \ref{section:social-planner}, to the social planner solution to determine the welfare losses induced by the firm's endogenous decisions. Finally, in Section \ref{section:opacity}, the role of opacity is isolated by solving the model under perfect information.

% Firm Problem

\section{Model} \label{section:structure}

\subsection{Production Technology and the Firm Objective}

\begin{figure}[H]
  \centering
  \begin{subfigure}{0.4\linewidth}
    \resizebox{\linewidth}{!}{\begin{tikzpicture}
	\begin{pgfonlayer}{nodelayer}
		\node [style=Dark blue firm] (01) at (-4, 3) {};
		\node [style=Dark blue firm] (02) at (-2, 3) {};
		\node [style=Dark blue firm] (03) at (0, 3) {};
		\node [style=Almost dark blue firm] (11) at (-3, 1) {};
		\node [style=Almost dark blue firm] (13) at (-1, 1) {};
	\end{pgfonlayer}
	\begin{pgfonlayer}{edgelayer}
		\draw [style=Supplies] (01) to (11);
		\draw [style=Supplies] (02) to (11);
		\draw [style=Supplies] (03) to (11);
		\node [draw, dotted,fit=(01) (02) (03), inner sep=6pt] (S11) {};
		\node [left, outer sep=10pt, inner sep=5pt] at (S11.west) {\large $\mathcal{S}_1$};

		\draw [style=Supplies] (02) to (13);
		\draw [style=Supplies] (03) to (13);
		\node [draw, dotted,fit=(02) (03), inner sep=4pt] (S22) {};
		\node [right, outer sep=10pt, inner sep=3pt] at (S22.east) {\large $\mathcal{S}_2$};
	\end{pgfonlayer}
\end{tikzpicture}}
  \end{subfigure} \hspace{2em}
  \begin{subfigure}{0.4\linewidth}
    \resizebox{\linewidth}{!}{\begin{tikzpicture}
	\begin{pgfonlayer}{nodelayer}
		\node [style=Dark blue firm] (01) at (-4, 3) {};
		\node [style=Not functional] (02) at (-2, 3) {};
		\node [style=Not functional] (03) at (0, 3) {};
		\node [style=Almost dark blue firm] (11) at (-3, 1) {};
		\node [style=Not functional] (13) at (-1, 1) {};
	\end{pgfonlayer}
	\begin{pgfonlayer}{edgelayer}
		\draw [style=Supplies] (01) to (11);
		\draw [style=Supplies] (02) to (11);
		\draw [style=Supplies] (03) to (11);
		\node [draw,dotted,fit=(01) (02) (03), inner sep=6pt] (S11) {};
		\node [left, outer sep=10pt, inner sep=5pt] at (S11.west) {\large $\mathcal{S}_1$};

		\draw [style=Supplies] (02) to (13);
		\draw [style=Supplies] (03) to (13);
		\node [draw,dotted,fit=(02) (03), inner sep=4pt] (S22) {};
		\node [right, outer sep=10pt, inner sep=3pt] at (S22.east) {\large $\mathcal{S}_2$};
	\end{pgfonlayer}
\end{tikzpicture}}
  \end{subfigure}
  \caption{The supply chain is depicted in the left panel. The left firm is sourcing its input good from all three suppliers, $\mathcal{S}_1$, while the right firm only from the latter two, $\mathcal{S}_2$. As a disruption occurs, some upstream firms are unable to supply the input good (white box). Unlike the left firm, the right firm is unable to source its inputs and is hence disrupted.}
  \label{fig:vertical}
\end{figure}
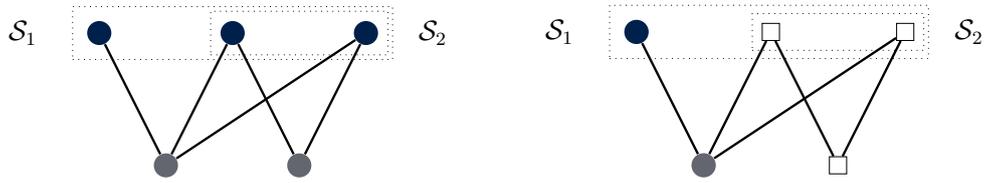

Consider a vertical economy producing $K + 1$ goods, indexed by $k \in \{0, 1, \ldots K\}$.  Each firm produces a single good and each good is produced by $m_k$ firms. Production of the \textit{basal good} $k = 0$ does not require any input, yet, it is at risk of random exogenous disruptions in the production process. A \textit{disrupted} basal firm is unable to deliver its good as input to downstream producers. Each downstream good $k > 0$ requires only good $k - 1$ as input. If a firm producing good $k$ is unable to source its input good $k - 1$, the firm is itself \textit{disrupted} and hence unable to deliver downstream. In other words, the $i$-th firm producing good $k$, indexed by $(k, i)$, is able to produce if at least one of its suppliers is able to deliver, namely, not all of its suppliers are disrupted. To avoid being disrupted, the firm chooses which firms to source from, among the producers of its input good. Letting $\D_{k}$ be the random set of disrupted firms in layer $k$ and $\S_{k, i}$ the set of suppliers of firm $(k, i)$, we can say that $(k, i) \in \D_{k}$ if and only if all of its suppliers $(k-1, j) \in \S_{k, i}$ are in $\D_{k - 1}$. I refer to the set of the firm's suppliers $\S_{k, i}$ as its \textit{sourcing strategy}. The disruption events are random and the probability that a firm is disrupted can be written as \begin{equation} \label{eq:functional_probability}
  P_{k, i} \coloneqq \P\big( (k, i) \in \D_k \big) = \P\big( \S_{k, i} \subset \D_{k-1} \big).
\end{equation} Figure \ref{fig:vertical} illustrates this mechanism. 

If a firm is not disrupted, it obtains a profit $\pi$. Implementing a given sourcing strategy costs the firm $C\big(\abs{\mathcal{S}_{k, i}}\big)$. The cost $C$ is assumed to be increasing in the number $\abs{\mathcal{S}_{k, i}}$ of suppliers. The problem of firm $(k, i)$ is then to maximise the expected profit\footnote{The expectation is taken over the random set $\mathcal{D}_{k - 1}$.} \begin{equation} \label{eq:profit-generic}
  \Pi_{k, i}(\S_{k, i}) = \Big(1 - \P\big( \S_{k, i} \subseteq \D_{k - 1} \big) \Big) \; \pi - C\big(\abs{\mathcal{S}_{k, i}}\big)
\end{equation} by picking a sourcing strategy $\S_{k, i}$. Before moving on with the solution of the model, it is useful to discuss the assumptions presented in this section. The production game is highly stylised: first, firms do not adjust prices but only quantities, such that failure to produce only arises in the case that no input is sourced; second, they are able to obtain profits by simply producing; third, contracting with new suppliers has a cost. There are both theoretical and empirical reasons behind these choices. Theoretically, a simpler model allows us to isolate the interplay between the variables of interest: correlation in the risk of suppliers, supply chain opacity, and the endogenous production network fragility. Empirically, these assumption capture well the rationale behind firms multisourcing. There is strong evidence that firms, first, when faced with supply chain shocks, adjust quantities rather than prices in the short run \citep{jiang_just--time_2022,lafrogne-joussier_supply_2022,di_giovanni_putting_2010,macchiavello_value_2015}, second, that production shutdowns can be extremely costly \citep{hameed_framework_2014, tan_general_1997}, and third, that fostering relationships with suppliers is costly, but important in guaranteeing operational performance \citep{cousins_implications_2006}. The model establishes a link between these issues faced by firms when choosing a sourcing strategy and the fragility of the production network.

\subsection{Opacity of the Supply Chain}

The supply chain is opaque: firms cannot observe the sourcing decisions of their potential suppliers before making their own. Furthermore, firms do not know how risky individual basal producers are, nor how their risk is correlated. Yet, firms know the distribution from which the probabilities of disruption in the basal layer are drawn. To motivate this assumption, recall the introductory example of hurricane Laura. A downstream firm producing PP, might not be able to trace back the production steps from its input to individual polymer producers in Louisiana or Texas, and, hence, the exact exposure of its production process to hurricanes. Yet, it can estimate the aggregate risk the polymer industry faces in the region. Given this information about the basal layer and their own depth $k$ in the production network, firms can derive the distribution of risk among their suppliers and make sourcing decisions based on it. By symmetry, the risk of two firms downstream sourcing from the same number of suppliers is ex-ante identical, albeit possibly correlated. The following two assumptions formalise this idea. Introduce \begin{equation}
  X_{k, j} \coloneqq \begin{cases}
    1 &\text{ if } (k, j) \text{ is disrupted and} \\
    0 & \text{ otherwise. }
  \end{cases}
\end{equation} \begin{assumption}\label{assumption:basal-exchangebility}
  Fix an arbitrary measure $\nu$ over $[0, 1]$. The probabilities $P_{0, j}$ of disruptions in the basal layer are sampled from $\nu$. I assume $\nu$ is observed by all firms, while $P_{0, j}$ are hidden.
\end{assumption} Going back to the example of polymer producers, under this assumption, downstream PP producers understand how hurricane risk can impact the production of their input good, via $\nu$, yet, they cannot estimate the risk that individual polymer producers face since they do not observe $P_{0, j}$.

\begin{assumption} \label{assumption:mixed-strategy}
  If there are multiple sourcing strategies that yield the same expected profit, the firm chooses one with equal probability. 
\end{assumption}

\begin{proposition} \label{proposition:exchangeability}
  Under these assumptions, in each downstream layer $k \geq 1$, disruption events \begin{equation*}
    X_{k, 1}, X_{k, 2}, X_{k, 3} \ldots X_{k, m_k},\end{equation*} are exchangeable, that is, their distribution is invariant under permutation.
\end{proposition}

Proposition \ref{proposition:exchangeability}, proven in Appendix \ref{appendix:proof-exchangeability}, asserts that, from the point of view of the firm, all suppliers are ex-ante identical, yet their risk might be correlated. Hence, the profit of the firm depends exclusively on \textit{how many} suppliers it chooses, rather than \textit{which} suppliers it chooses. Then, a firm producing good $k$ can first infer the distribution of the number $D_{k - 1} \coloneqq \abs{\D_{k - 1}}$ of disrupted firms among its potential suppliers and then choose the optimal number $s_{k, i} \coloneqq \abs{\S_{k - 1, i}}$ of firms from which to source its input good. Furthermore, by symmetry, all firms in layer $k$ choose the same number $s_k$ of sources, that is, \begin{equation}
  s_{k, i} = s_k \text{  for all  } i.
\end{equation} As a result, the sourcing strategies $\S_{k, i}$ and $\S_{k, j}$ of any two firms $i$ and $j$ are such that their disruption probabilities $P_{k, i}$ and $P_{k, j}$ are identically distributed\footnote{This approach is widely used in the study of random graphs, see, for example, \cite{kallenberg_probabilistic_2005,diaconis_graph_2007}}.

\section{Disruptions Propagation} \label{section:propagation}

Building on the mechanisms behind firms disruptions introduced above, this section studies how these disruptions propagate through the supply chain. To do so, I consider the case in which the number $m_k$ of firms in each layer $k$ grows large. To study the limit, it is first necessary to characterise how the sourcing relations $\S_{k, j}$ form as the number of firms in each layer increases.

\begin{assumption} \label{assumption:limiting}
  As a new firm is introduced in layer $k$, it starts establishing relations with its $s_k$ suppliers. As soon as it pairs with a supplier, a new firm is introduced among the producers of its input good $k - 1$, which, in turn, selects its sources. This procedure continues recursively until all firms realised their sourcing strategy $s_k$.
\end{assumption}

Indexing by $n$ the $n$-th step of this procedure, this section focuses on the limit as $n \to \infty$. Every new firm introduced in the basal layer has a disruption probability that is $\nu$-distributed, hence, the new firm is ex-ante identical to existing firms. This ensures that, as $n \to \infty$, Assumption \ref{assumption:basal-exchangebility} is satisfied and the downstream sourcing decisions $s_1, s_2 \ldots$ are unaffected. This, allows us to simply consider the problem of the representative firm in layer $k$. 

To analytically characterise the disruption propagation through the production network, the only missing piece is the distribution $\nu$ of the disruption probabilities in the basal layer. As mentioned in the previous section, I assume that basal firms fail with a not necessarily independent probability $P_0$. We can model this by assuming that $P_0$ follows a $\Beta$ distribution. \begin{assumption} \label{assumption:basal-distribution}
  The probability of a disruption in the basal layer follows \begin{equation}
    P_{0} \sim \nu_0 \equiv \Beta \text{ for all } j.
  \end{equation}
\end{assumption} The $\Beta$ distribution allows to flexibly model shocks that might happen due to spacial or technological proximity of basal producers, which cannot be diversified. Consider, for example, how oil extraction plants must be located nearby oil reserves and are hence all subject to correlated weather shocks that might force them to shut down. In this case, despite the small expected probability that an individual firm is disrupted, as a hurricane is a rare occurrence, disruptions are highly correlated, as when a hurricane occurs most of them are disrupted. To keep track of the expected disruption probability and the correlation of risk through the layers, I introduce the following alternative parametrisation of the Beta distribution.

\begin{definition} \label{definition:parametrisation}
  Let $\mu$ and $\rho$ be respectively the mean and the overdispersion of a $\Beta$ distribution with shape parameters $\alpha$ and $\beta$, defined by \begin{equation}
    \mu \coloneqq \frac{\beta}{\alpha + \beta} \text{ and } \rho \coloneqq \frac{1}{1 + \alpha + \beta}.
  \end{equation} I write $P \sim \Beta(\mu, \rho)$.
\end{definition}

Given Assumption \ref{assumption:basal-distribution}, the following result links the probabilities of experiencing a disruption from upstream suppliers of $k$ to downstream producers $k + 1$. 

\begin{definition} \label{definition:BetaPower}
  A random variable $Y$ follows a $\BP$ distribution, with mean $\mu$, overdispersion $\rho$, and power $s$ if it can be written as $Y = X^s$ where $X$ follows a $\Beta$ distribution with mean $\mu$ and overdispersion $\rho$.
\end{definition}

\begin{proposition} \label{proposition:PtoP}
  If the disruption probability $P_k$ among suppliers of good $k$ follows a $\BP$ distribution, so does the downstream probability $P_{k + 1}$. 
\end{proposition}

\begin{figure}[H]
  \centering
  \input{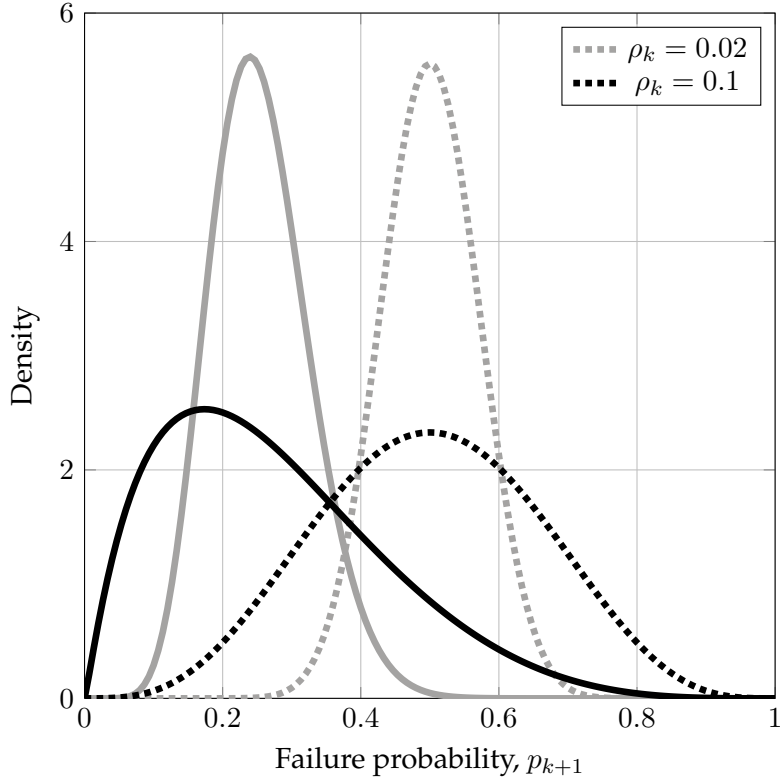}
  \caption{Distribution of disruption probabilities of downstream firms for different levels of upstream correlation $\rho_k$, in the cases of single sourcing (dotted) and multisourcing (solid). In both cases $\mu_k = 1/2$.}
  \label{fig:distribution-illustration}
\end{figure}

The proof is provided in Appendix \ref{appendix:PtoP}. Proposition \ref{proposition:PtoP} guarantees that the distribution of disrupted firms will remain in the same distribution family as risk amplifies through the production network. This result allows us to describe disruption propagation in the supply chain by mapping the evolution of the parameters $\mu_k$ and $\rho_k$ through the layers. Furthermore, it allows firms to estimate $\mu_k$ and $\rho_k$ and use this to determine the optimal sourcing strategy $s_{k + 1}$. It is useful at this point to give an interpretation of $\mu_k$ and $\rho_k$ in the context of our model. The parameter $\mu_k$ is the average failure probability of firms in layer $k$. The parameter $\rho_k$ tracks the degree of correlation in the disruption of firms operating in layer $k$. I illustrate this in Figure \ref{fig:distribution-illustration}. This figure shows the distribution of the disruption probability $P_{k + 1}$ among downstream firms in the case the firm has a single supplier (dotted lines) or two suppliers (solid line). For low overdispersion, $\rho_k = 0.01$, the suppliers' disruptions are weakly correlated and the downstream disruption probability is concentrated around the average $\mu_k$. If firms contract an additional supplier, the distribution of failures decreases and remains concentrated around the average. As $\rho_k$ increases, the suppliers' disruption events become more correlated and the downstream disruption probabilities become fat-tailed, that is, a significant fraction of firms is likely to be disrupted and, as a consequence, diversification is ineffective. If firms contract an additional supplier the average disruption probability decreases, but a large disruption probability remains. 

Having established the link between the disruptions of layer $k$ to layer $k + 1$, I now turn to the analysis of how these propagate through the whole supply chain, before studying how firms make decisions endogenously. The following result recursively connects downstream distributions with upstream sourcing decisions and initial conditions.

\begin{proposition} \label{proposition:risk-distribution}
  The average disruption probability between one layer $k$ and the next $k + 1$ depends on the sourcing strategy $s_{k + 1}$ via \begin{equation}
    \mu_{k + 1} = \begin{cases} 
      \eta(s_{k + 1}, S_k) \; \mu_{k} &\text{ if } s_{k + 1} > 0, \\
      1 &\text{ otherwise, }
    \end{cases}
  \end{equation} where $S_k \coloneqq \prod^k_{j = 1} s_j$ is the diversification level up to layer $k$ and $\eta$ is the ``risk reduction factor'', which is given by \begin{equation} \label{eq:eta}
    \begin{split}
      \eta(s_{k + 1}, S_k) &= \left(\mu_0 \frac{1 - \rho_0}{\rho_0} + S_k\right)^{\overline{S_k s_{k + 1}}} \Bigg/ \left(\frac{1 - \rho_0}{\rho_0} + S_k\right)^{\overline{S_k s_{k + 1}}} \\
      &= \left(\frac{\mu_0 \; \frac{\rho_0}{1 - \rho_0} + S_k}{\frac{\rho_0}{1 - \rho_0} + S_k} \right)\left(\frac{\mu_0 \; \frac{\rho_0}{1 - \rho_0} + S_k + 1}{\frac{\rho_0}{1 - \rho_0} + S_k + 1} \right) \dots  \left(\frac{\mu_0 \; \frac{\rho_0}{1 - \rho_0} + S_k s_{k + 1} - 1}{\frac{\rho_0}{1 - \rho_0} + S_k s_{k + 1} - 1} \right).
    \end{split}
  \end{equation}
\end{proposition}

This is proven in Appendix \ref{appendix:derivations}. The risk reduction factor $\eta(s_{k + 1}, S_k)$ governs how the firm's choice $s_{k + 1}$, the choices along the firm's production chain $S_k$, and the basal conditions $\mu_0, \rho_0$ affect the expected number of disruptions downstream. This interplay is illustrated in the following figures. 

Figure \ref{fig:risk-mapping} shows how the risk reduction factor varies with basal correlation $\rho_0$ for different sourcing strategies $s_k$, fixing the upstream diversification of $S_k = 2$. If correlation $\rho_0$ in the basal layer grows, to obtain a given level of risk reduction $\eta$, firms producing good $k$ needs to source more suppliers. If $\rho_0 \to 1$, diversification becomes impossible, as $\eta \to 1$ and $\mu_{k + 1} \to  \mu_k$ for any sourcing strategy $s_{k + 1}$. 

\begin{figure}[H]
  \centering
  \input{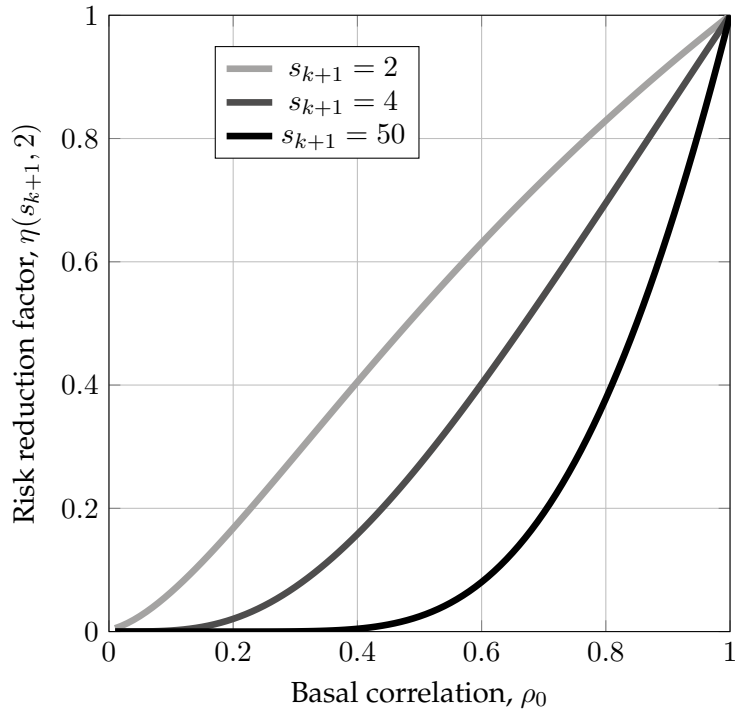}
  \caption{Risk reduction factor $\mu_{k + 1} / \mu_{k}$ at different basal overdispersion levels $\rho_0$ and for different sourcing strategies $s_{k + 1}$. $S_k = 2$.}
  \label{fig:risk-mapping}
\end{figure}

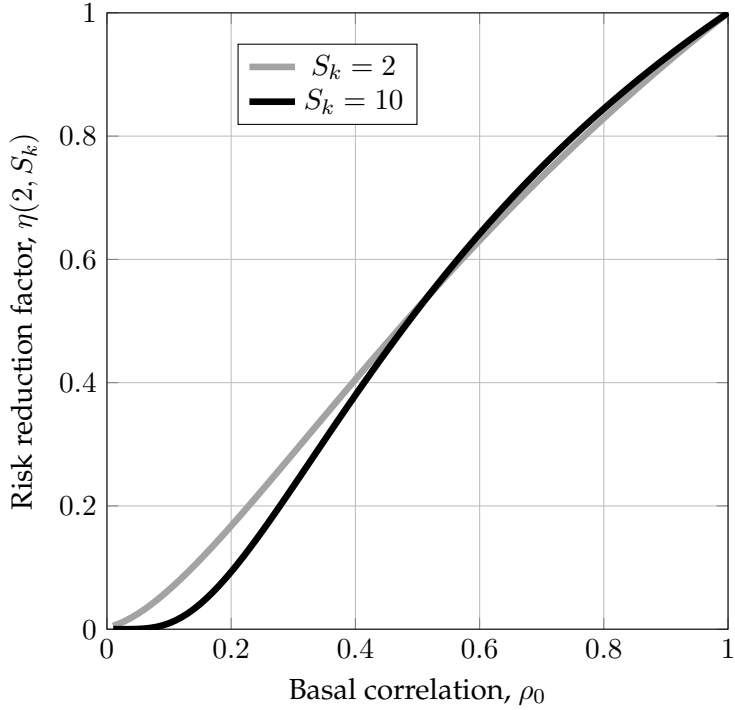
\begin{figure}[H]
  \centering
  % Recommended preamble:
\begin{tikzpicture}
\begin{axis}[width={0.64\textwidth}, height={0.64\textwidth}, xmin={0}, xmax={1}, ymin={0}, ymax={1}, grid={both}, no markers, xtick={0.0,0.2,0.4,0.6,0.8,1.0}, xlabel={Basal correlation, $\rho_0$}, ylabel={Risk reduction factor, $\eta(2, S_k)$}, legend style={at = {(0.5, 0.95)}}]
    \addplot[color={rgb,1:red,0.642;green,0.6405;blue,0.6372}, line width={2.5}]
        coordinates {

            (0.01,0.00536327897495632)
            (0.02,0.009144984917043741)
            (0.03,0.013763738779996338)
            (0.04,0.019145299145299145)
            (0.05,0.025221861471861475)
            (0.06,0.031931435309973055)
            (0.07,0.03921728971962618)
            (0.08,0.047027458492975745)
            (0.09,0.05531429793189238)
            (0.1,0.06403409090909092)
            (0.11,0.07314669177374096)
            (0.12,0.08261520737327188)
            (0.13,0.09240571007163931)
            (0.14,0.10248697916666666)
            (0.15,0.11283026755852839)
            (0.16,0.12340909090909093)
            (0.17,0.1341990368669473)
            (0.18,0.14517759222333)
            (0.19,0.15632398611618561)
            (0.2,0.16761904761904764)
            (0.21,0.17904507624257943)
            (0.22,0.19058572404371588)
            (0.23,0.20222588818353934)
            (0.24,0.21395161290322576)
            (0.25,0.22575)
            (0.26,0.237609126984127)
            (0.27,0.2495179721852951)
            (0.28,0.2614663461538462)
            (0.29,0.27344482877048365)
            (0.3,0.28544471153846146)
            (0.31,0.29745794458580727)
            (0.32,0.3094770879526977)
            (0.33,0.3214952667814115)
            (0.34,0.333506130063966)
            (0.35,0.34550381263616553)
            (0.36,0.35748290013679895)
            (0.37,0.3694383966775737)
            (0.38,0.38136569499341233)
            (0.39,0.39326054886427947)
            (0.4,0.40511904761904766)
            (0.41,0.4169375925492945)
            (0.42,0.42871287507654615)
            (0.43,0.4404418565305661)
            (0.44,0.45212174940898325)
            (0.45,0.4637499999999999)
            (0.46,0.47532427226027407)
            (0.47,0.4868424328494284)
            (0.48,0.49830253723110857)
            (0.49,0.509702816758186)
            (0.5,0.5210416666666665)
            (0.51,0.5323176349091862)
            (0.52,0.5435294117647058)
            (0.53,0.5546758201662542)
            (0.54,0.5657558066933068)
            (0.55,0.5767684331797237)
            (0.56,0.5877128688921142)
            (0.57,0.5985883832370974)
            (0.58,0.6093943389592121)
            (0.59,0.6201301857942415)
            (0.6,0.6307954545454544)
            (0.61,0.6413897515527951)
            (0.62,0.651912753527337)
            (0.63,0.6623642027254464)
            (0.64,0.6727439024390245)
            (0.65,0.6830517127799737)
            (0.66,0.6932875467386789)
            (0.67,0.7034513664977737)
            (0.68,0.7135431799838579)
            (0.69,0.723563037641092)
            (0.7,0.7335110294117645)
            (0.71,0.7433872819100092)
            (0.72,0.7531919557758292)
            (0.73,0.7629252431975186)
            (0.74,0.772587365591398)
            (0.75,0.7821785714285713)
            (0.76,0.7916991341991341)
            (0.77,0.8011493505049156)
            (0.78,0.8105295382724719)
            (0.79,0.819840035078602)
            (0.8,0.8290811965811966)
            (0.81,0.8382533950487115)
            (0.82,0.8473570179820177)
            (0.83,0.8563924668227947)
            (0.84,0.8653601557430239)
            (0.85,0.8742605105105103)
            (0.86,0.8830939674256799)
            (0.87,0.8918609723252275)
            (0.88,0.9005619796484737)
            (0.89,0.9091974515625595)
            (0.9,0.9177678571428571)
            (0.91,0.9262736716052131)
            (0.92,0.9347153755868546)
            (0.93,0.9430934544729884)
            (0.94,0.951408397766323)
            (0.95,0.9596606984969052)
            (0.96,0.9678508526698351)
            (0.97,0.9759793587485757)
            (0.98,0.9840467171717174)
            (0.99,0.9920534299011838)
            (1.0,1.0)
        }
        ;
    \addlegendentry {$S_{k} = 2$}
    \addplot[color={rgb,1:red,0.0;green,0.0;blue,0.0}, line width={2.5}]
        coordinates {

            (0.01,1.9603234329285155e-8)
            (0.02,1.601030360847433e-6)
            (0.03,2.015539296708888e-5)
            (0.04,0.0001100500338742874)
            (0.05,0.00037795013328007193)
            (0.06,0.0009716889216339247)
            (0.07,0.0020554804483275)
            (0.08,0.00378551439697722)
            (0.09,0.006293019668178768)
            (0.1,0.009675808211664865)
            (0.11,0.013996487377674394)
            (0.12,0.019284925717716908)
            (0.13,0.025542929403896207)
            (0.14,0.03274970538601196)
            (0.15,0.04086724408844108)
            (0.16,0.04984516284155066)
            (0.17,0.059624817704017834)
            (0.18,0.07014264964984308)
            (0.19,0.08133281594499098)
            (0.2,0.09312919623234779)
            (0.21,0.1054668744650663)
            (0.22,0.1182831947757203)
            (0.23,0.13151847926018437)
            (0.24,0.14511648293140675)
            (0.25,0.15902464819112144)
            (0.26,0.17319420929099)
            (0.27,0.18758018691417733)
            (0.28,0.20214130431788488)
            (0.29,0.2168398493496992)
            (0.3,0.23164150090826488)
            (0.31,0.2465151338558416)
            (0.32,0.2614326128047935)
            (0.33,0.2763685824093761)
            (0.34,0.2913002596408696)
            (0.35,0.30620723187635684)
            (0.36,0.321071263381274)
            (0.37,0.33587611182607974)
            (0.38,0.3506073557784099)
            (0.39,0.3652522335987834)
            (0.4,0.3797994937970065)
            (0.41,0.3942392566440498)
            (0.42,0.40856288665393653)
            (0.43,0.4227628754317956)
            (0.44,0.4368327343118291)
            (0.45,0.45076689617051857)
            (0.46,0.4645606257866939)
            (0.47,0.47820993812399754)
            (0.48,0.4917115239275178)
            (0.49,0.5050626820509487)
            (0.5,0.5182612579604132)
            (0.51,0.5313055878940486)
            (0.52,0.5441944481907468)
            (0.53,0.5569270093360708)
            (0.54,0.5695027943073917)
            (0.55,0.5819216408331757)
            (0.56,0.5941836672127465)
            (0.57,0.6062892413724355)
            (0.58,0.618238952861852)
            (0.59,0.6300335875197729)
            (0.6,0.641674104563158)
            (0.61,0.6531616158748215)
            (0.62,0.6644973672856309)
            (0.63,0.675682721665661)
            (0.64,0.6867191436558044)
            (0.65,0.6976081858868236)
            (0.66,0.7083514765470073)
            (0.67,0.7189507081724977)
            (0.68,0.7294076275460114)
            (0.69,0.7397240266004088)
            (0.7,0.7499017342331227)
            (0.71,0.7599426089463117)
            (0.72,0.7698485322354933)
            (0.73,0.7796214026566396)
            (0.74,0.7892631305082413)
            (0.75,0.7987756330707597)
            (0.76,0.808160830351255)
            (0.77,0.8174206412858124)
            (0.78,0.8265569803568058)
            (0.79,0.8355717545860317)
            (0.8,0.8444668608683118)
            (0.81,0.8532441836134965)
            (0.82,0.8619055926677003)
            (0.83,0.8704529414873468)
            (0.84,0.8788880655420014)
            (0.85,0.8872127809241676)
            (0.86,0.8954288831462757)
            (0.87,0.9035381461068529)
            (0.88,0.9115423212095508)
            (0.89,0.9194431366201711)
            (0.9,0.9272422966482311)
            (0.91,0.9349414812407866)
            (0.92,0.942542345577409)
            (0.93,0.9500465197561774)
            (0.94,0.9574556085614957)
            (0.95,0.9647711913053985)
            (0.96,0.971994821734743)
            (0.97,0.9791280279973916)
            (0.98,0.9861723126611157)
            (0.99,0.9931291527795528)
            (1.0,1.0)
        }
        ;
    \addlegendentry {$S_{k} = 10$}
\end{axis}
\end{tikzpicture}
  \caption{Risk reduction factor $\mu_{k + 1} / \mu_{k}$ at different basal overdispersion levels $\rho_0$ and for different upstream diversification levels strategies $S_{k}$. $s_{k + 1} = 2.$}
  \label{fig:upstream-risk-mapping}
\end{figure}

As the above, Figure \ref{fig:upstream-risk-mapping} shows the response of the risk reduction factors to different levels of basal correlation, but instead of varying the strategy $s_k$ of the firm, it varies the level of upstream diversification $S_k$. For low levels of basal correlation $\rho_0$, more upstream diversification $S_k$ allows downstream producers to achieve lower risk with fewer suppliers. Yet, there is a level of basal correlation after which more diversification is detrimental for the downstream firm, as this high upstream diversification simply exacerbates tail-risk. This represents a crucial externality the upstream suppliers impose on downstream producers. For low level of correlation, sourcing downstream represents a positive externality downstream. This externality shrinks as correlation increases, until it becomes a negative externality. Section \ref{section:social-planner} explores the welfare consequences of this mechanism.

\section{Firm Optimal Diversification and Competitive Equilibrium} \label{section:optimum}

The mechanics of disruption propagation, derived in the previous section, determines the firm's desired sourcing strategy and, as a consequence, their optimal sourcing strategy. This section derives such optimal strategies. Importantly, due to Proposition \ref{proposition:exchangeability}, all firms in a given layer are identical before the shock and so is their optimisation problem. We can hence focus of the problem of the representative firm in layer $k + 1$: to choose how many suppliers in layer $k$ to source from, based on the inferred distribution of their probability of experiencing a disruption event. This, in turn, is fully determined by average disruption probability $\mu_0$, the correlation $\rho_0$ of the disruptions, and the sourcing strategies $\{s_1, s_2, \ldots s_k\}$ of the representative firms upstream. Henceforth, I assume firms face quadratic costs of sourcing, with cost parameter $c$, such that expected profits (\ref{eq:profit-generic}) can be written as \begin{equation} \label{eq:profits}
  \Pi_k(s) = \left(1 - \E_s\big[P_k\big]\right) \pi - \frac{c}{2} s^2,
\end{equation} The optimisation problem of the firm is to then choose the optimal sourcing strategy \begin{equation} \label{eq:problem}
  s_{k} = \arg\max_{s \in \{0, 1, 2, \ldots\}} \Pi_k(s).
\end{equation}

\subsection{Limit Case: Uncorrelated Disruptions}

Before turning towards the general framework, I first analyse a limit case in which suppliers' risk is not correlated, that is $\rho_0 \to 0$. This limit case gives a useful interpretation of the incentives behind multisourcing and allows us to establish a benchmark against which to study the introduction of correlated shocks. 

\begin{proposition} \label{proposition:exogenous-law-of-motion:no-corr}
  If risk among basal firms is uncorrelated, that is $\rho_0 \to 0$, disruption events in layer $k$ are independent and happen with probability \begin{equation} \label{eq:exogenous-law-of-motion:no-corr} 
    \mu_{k + 1} = \mu_{k}^{s_{k + 1}}.
  \end{equation}
\end{proposition}

\begin{proof}
  Follows immediately from $P_k \to \mu_k$ as $\rho_0 \to 0$.
\end{proof}

As in this case $\E_{s}[P_{k + 1}] = \mu_k^{s}$, profits (\ref{eq:profits}) are given by $\Pi_{k + 1}(s) = \big(1 - \mu_k^s\big) \pi - \frac{c}{2} s^2$. Using this, we can derive the optimal sourcing strategy $s_k$ of a firm producing good $k$. A firm with $s$ suppliers, contracts an extra one only if doing so yields a positive marginal profit \begin{equation}
  \begin{split}
    \Delta \Pi_{k + 1}(s) &\coloneqq \Pi_{k + 1}(s + 1) - \Pi_{k + 1}(s)\\
    &= \mu_k^s (1 - \mu_k) \pi - c \left(s + \frac{1}{2}\right).
  \end{split}
\end{equation} It is easy to check that $\Delta \Pi_{k + 1}(s)$ is strictly decreasing and that it has a unique root. Hence, the optimal number of suppliers $s_{k + 1}$ is the smallest integer $s$ for which the expected marginal profit is negative, that is $\Delta \Pi_{k + 1}(s) < 0$.

\begin{definition} \label{definition:no-corr:incentives}
  Let $\tilde{s}_{k + 1} $ be the unique real root of $\Delta \Pi_{k + 1}$. I refer to this quantity as the ``desired sourcing strategy'' of the firm.
\end{definition}

The optimal sourcing strategy is then given by \begin{equation}
  s_{k + 1} = \begin{cases}
    \ceil{\tilde{s}_{k + 1}} &\text{if } \tilde{s}_{k + 1} > 0 \text{ and}\\
    0 &\text{otherwise.}
  \end{cases}
\end{equation}

\begin{proposition} \label{proposition:no-sourcing:no-corr}
  Introduce the threshold \begin{equation}
    \mu^0 \coloneqq 1 - rc
  \end{equation} where $rc \coloneqq \frac{c/2}{\pi}$ is the real marginal costs of an additional supplier. If the average disruption probability $\mu_k$ is larger than $\mu_0$, the downstream firm does not source any inputs, that is $s_{k + 1} = 0$.
\end{proposition}

\begin{proof}
  Suppose a firm optimally does not source any inputs. This implies that the marginal benefit of adding the first supplier is negative, namely $\Delta \Pi_{k + 1}(0) < 0$, which yields the desired inequality.
\end{proof}

\begin{figure}[H]
  \centering
  \input{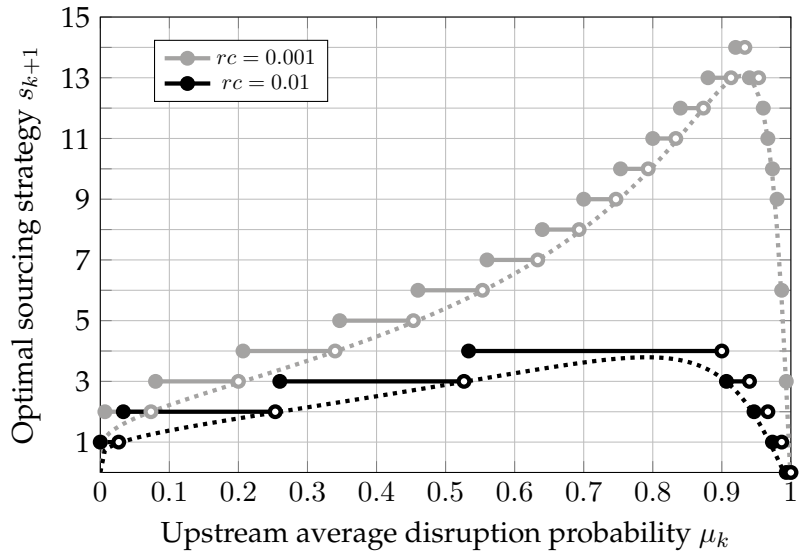}
  \caption{The desired $\tilde{s}_{k+1}$ (dotted) and optimal $s_{k + 1}$ sourcing strategy (solid) as a function of the upstream average disruption probability $\mu_k$}
  \label{fig:s-no-corr}
\end{figure}

As expected, the desired $\tilde{s}_{k + 1}$ and the optimal sourcing strategy $s_{k + 1}$ are determined by the upstream average disruption probability $\mu_k$ and the real marginal costs of contracting a new supplier $rc$. Figure \ref{fig:s-no-corr} illustrates the effect these two conditions have on the optimal sourcing strategy. First, higher real marginal costs $rc$ reduce the firm's number of sources. Second, as the the upstream average disruption probability $\mu_k$ increases, initially the firm seeks higher diversification, until a level above which the desired sourcing strategy start decreasing steeply. 

Having studied how risk affects the firm's optimal sourcing, I now look at the opposite channel, that is, how does the firm sourcing strategy affect risk propagation. To do so, we think of the average disruption probability \begin{equation} \label{eq:law-of-motion:no-corr}
  \mu_{k + 1} = \mu_k^{s_{k + 1}(\mu_k)}
\end{equation} from suppliers to downstream producers as a dynamical system, not in time but in layers $k \in \{0, 1, 2 \ldots \}$ of the supply chain. Given a \textit{basal condition} $\mu_0$, a fixed point $\bar{\mu}$ of the map (\ref{eq:law-of-motion:no-corr}) is then a level of disruption probability $\bar{\mu}$ such that all firms downstream of a layer $\overline{k}$ single-source, namely $s_{l} = 1$ for all $l \geq \overline{k}$, and hence all share the same disruption probability $\mu_l \equiv \bar{\mu}$. When looking at the production network through this lens, a natural question arises: which basal levels of disruption probabilities $\mu_0$ are not endogenously diversified by the production network, that is $\bar{\mu} \geq \mu_0$? To answer this, first I characterise the downstream disruption probability $\bar{\mu}$.

\begin{proposition} \label{proposition:steady-state:no-corr}
  The downstream disruption probability $\bar{\mu}$ satisfies \begin{equation}
    \bar{\mu} \; (1 - \bar{\mu}) \leq 3rc.
  \end{equation}
\end{proposition}

\begin{proof}
  A steady state is attained at level $\bar{k}$ if $s_{\bar{k}} = 1$. This implies that the marginal benefit of multisourcing is negative $\Delta\Pi_{\bar{k}}(1) \leq 0$. This yields the desired inequality.
\end{proof}

\begin{corollary} \label{corollary:basin:no-corr}
  Introduce the critical threshold \begin{equation}
    \mu^c \coloneqq \frac{1}{2} + \sqrt{\frac{1}{4} - 3rc}.
  \end{equation} If $\mu_0 > \mu^c$, the endogenous supply chain is unable to diversify risk, that is $\bar{\mu} \geq \mu_0$.
\end{corollary}

\begin{wrapfigure}{r}{0.4\textwidth}
  % Recommended preamble:
\begin{tikzpicture}
\begin{axis}[width={\linewidth}, height={\linewidth}, xlabel={$rc$}, ylabel={$\mu$}, grid={both}, xtick={0.0,0.02,0.04,0.06,0.08}, xticklabels={0.0,0.02,0.04,0.06,0.08}, xmin={0}, xmax={0.08}, ymin={0.5}, ymax={1}, ytick={0.6,0.7,0.8,0.9,1.0}, scaled x ticks={false}, legend style={at = {(0.4, 0.4)}}]
    \addplot[color={rgb,1:red,0.0;green,0.0;blue,0.0}, ultra thick, forget plot]
        coordinates {
            (0.0,1.0)
            (0.01,0.99)
            (0.02,0.98)
            (0.03,0.97)
            (0.04,0.96)
            (0.05,0.95)
            (0.06,0.94)
            (0.07,0.9299999999999999)
            (0.08,0.92)
        }
        ;
    \addplot[color={rgb,1:red,0.0;green,0.0;blue,0.0}, ultra thick, only marks]
        coordinates {
            (0.0,1.0)
            (0.01,0.99)
            (0.02,0.98)
            (0.03,0.97)
            (0.04,0.96)
            (0.05,0.95)
            (0.06,0.94)
            (0.07,0.9299999999999999)
            (0.08,0.92)
        }
        ;
    \addlegendentry {$\mu^{0}$}
    \addplot[color={rgb,1:red,0.642;green,0.6405;blue,0.6372}, ultra thick, only marks, line width={2.5}]
        coordinates {
            (0.0,1.0)
            (0.01,0.969041575982343)
            (0.02,0.9358898943540673)
            (0.03,0.9)
            (0.04,0.860555127546399)
            (0.05,0.8162277660168379)
            (0.06,0.764575131106459)
            (0.07,0.7)
            (0.08,0.6000000000000001)
        }
        ;
    \addplot[color={rgb,1:red,0.642;green,0.6405;blue,0.6372}, ultra thick, forget plot, line width={2.5}]
        coordinates {
            (0.0,1.0)
            (0.01,0.969041575982343)
            (0.02,0.9358898943540673)
            (0.03,0.9)
            (0.04,0.860555127546399)
            (0.05,0.8162277660168379)
            (0.06,0.764575131106459)
            (0.07,0.7)
            (0.08,0.6000000000000001)
        }
        ;
    \addlegendentry {$\mu^c$}
\end{axis}
\end{tikzpicture}
  \caption{}
  \label{fig:threshold}
\end{wrapfigure}
This result links the firm real marginal costs of sourcing $rc$ and the production network risk. As relative marginal costs increase, the capacity of the production network to endogenously diversify decreases and firms underdiversification yields endogenous fragility. Notice that, comparing the threshold $\mu^c$ of endogenous diversification with the threshold $\mu^{0}$ of firm shutdown, illustrated in Figure \ref{fig:threshold}, for some levels of basal probability of disruption $\mu_0$, despite no firm shutting down production $\mu_0 < \mu^{0}$, the production network as a whole is still unable to endogenously diversify risk $\mu_0 > \mu^c$. This is true even in this special case, where firms risk is uncorrelated. In the next section, I introduce correlation risk $\rho > 0$ and investigate how doing so changes the dynamics illustrated here.

\subsection{Optimal Sourcing with Correlated Distributions}

\begin{figure}[H]
  \centering
  \input{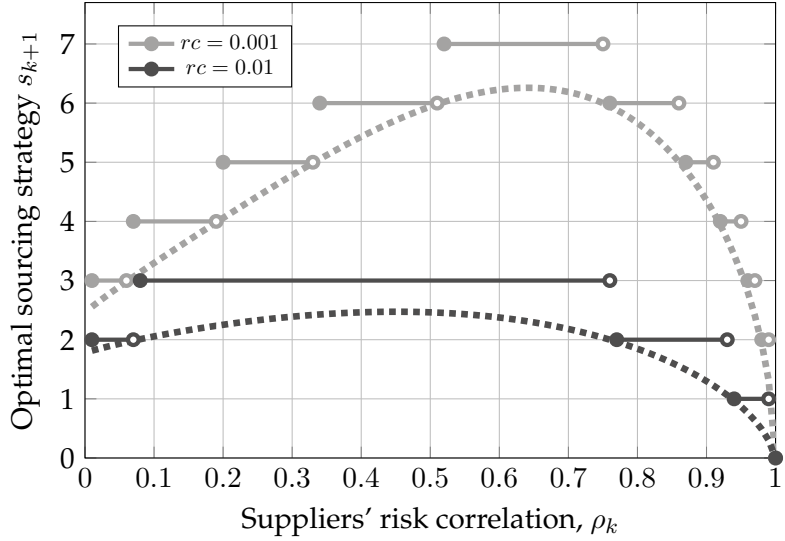}
  \caption{The desired $\tilde{s}_{k+1}$ (dotted) and optimal $s_{k + 1}$ sourcing strategy (solid) as a function of the upstream correlation $\rho_k$}
  \label{fig:opt-s}
\end{figure}

If disruption events are not independent, that is $\rho_0 > 0$, risk among suppliers throughout the production network is correlated, which affects the firm's optimisation incentives. In this case, the problem of a firm in layer $k + 1$ is still to choose the number of suppliers $s_{k + 1} \in \{0,1,2 \ldots\}$ that maximises the profits $\Pi$, but, by Proposition \ref{proposition:risk-distribution}, the firm's disruption probability is given by the average disruption probability of its suppliers $\mu_{k}$ multiplied by a factor $\eta(s_{k + 1}, S_k)$ which depends on the upstream diversification $S_k$. As in the limit case analysed in the previous section, the firm will increase diversification as long as the expected increase in profits obtained by adding an additional supplier outweighs the costs of contracting that additional supplier. These expected marginal profits are given by \begin{equation}
  \Delta \Pi_{k + 1}(s_{k + 1}) = \Big(\eta(s_{k + 1}, S_k) - \eta(s_{k + 1} + 1, S_k)\Big) \; \mu_k \pi - c \left(s_{k + 1} + \frac{1}{2}\right).
\end{equation} The characterisation of the optimal sourcing strategy is analogous to the limit case without correlation discussed above. $\tilde{s}_{k + 1} \in \mathbb{R}$ is the desired sourcing strategy for which the marginal benefits and marginal costs of diversification are equal, such that $\Delta\Pi_{k + 1}(\tilde{s}_{k + 1}) = 0$. As the marginal profits are strictly decreasing in the number of suppliers (see Appendix \ref{appendix:general-case-proofs}), the firm will, as in the limit case, choose its optimal sourcing strategy as $s_{k + 1} = \ceil{\tilde{s}_{k + 1}}$ if $\tilde{s}_{k + 1} > 0$ and chooses not to source any inputs otherwise. Figure \ref{fig:opt-s} illustrates how the optimal sourcing strategy $s_{k + 1}$ changes with upstream correlation $\rho_k$ for different levels of relative costs $rc$. As upstream correlation increases, the firm increases its sources to diversify risk. Yet, for large levels of correlation, the disruption of an additional source of the input good is likely correlated to a disruption among the firm's existing suppliers, which reduces the firm's incentive to multisource. As disruptions among suppliers become perfectly correlated, $\rho_k \to 1$, the firm sources from a single supplier.

We have the following formal result, proved in Appendix \ref{appendix:general-case-proofs}. \begin{proposition} \label{proposition:concavity-of-incentives}
  The desired optimal sourcing $\tilde{s}_{k+1}$ is concave in the overdispersion among suppliers $\rho_k$.
\end{proposition} \begin{corollary} \label{corollary:concavity-of-opt}
  There is a level of correlation $\rho^c$ such that $s_{k + 1}$ is (weakly) increasing in $[0, \rho^c)$ and (weakly) decreasing in $[\rho^c, 1]$. 
\end{corollary} 

This result establishes the trade-off the firm faces when supplier risk is correlated: on the one hand, more correlation increases the benefits of diversification, and on the other hand, it makes such diversification less effective. 

\begin{figure}[H]
  \centering
  \input{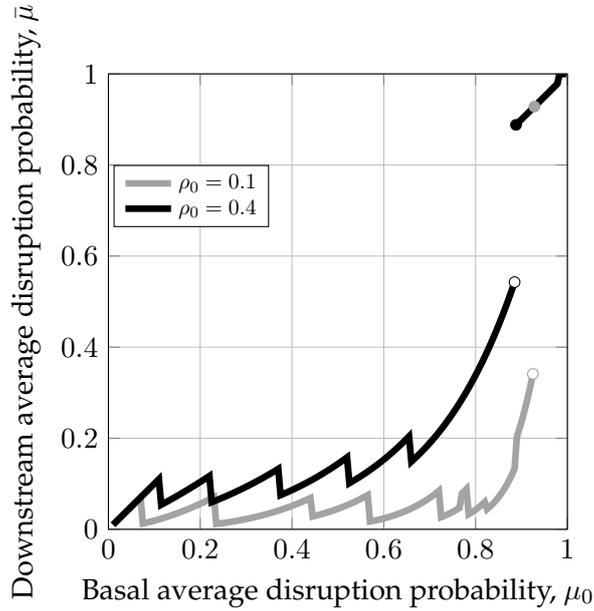}
  \caption{Downstream average disruption probability as a function of basal average disruption probability for low and high basal levels of correlation $\rho_0$}
  \label{fig:limit}
\end{figure}

To study the ramifications this endogenous channel has on the supply chain formation and its fragility, in the following I analyse the propagation of risk through the layers. As above, we can view the mapping of the probability of disruption between layers (\ref{proposition:risk-distribution}) as a dynamical system through the layers of the production network. A steady state of the dynamical system is then an average disruption probability $\bar{\mu} \coloneqq \mu_{\bar{k}}$ in some layer $\bar{k}$ such that all downstream layers $l \geq \bar{k}$ have the same average disruption probability $\mu_l \equiv \bar{\mu}$. This can occur in two cases. Either the firm in layer $\bar{k}$ does not source, that is $s_{\bar{k}} = 0$, or it single sources, that is $s_{\bar{k}} = 1$. The former case is trivial: the production network shuts down and all downstream firms do not produce, such that $\bar{\mu} = 1$. In the latter case, by single sourcing, the average disruption probability in layer $\bar{k}$ is the average disruption probability among the suppliers $\bar{k} - 1$, as the risk reduction factor $\eta(s_k, S_{k - 1}) = 1$ if $s_{k} = 1$. Because the layers are symmetric, the firms in the downstream layer $\bar{k} + 1$ face then the same problem as those in layer $\bar{k}$, such that they endogenously single source, that is $s_{\bar{k} + 1} = 1$. Inductively, this holds true for all $l \geq \bar{k}$, hence $\mu_{l} \equiv \bar{\mu}$. Hereafter, I refer to the situation in which the downstream average disruption probability is greater than the basal one, that is $\bar{\mu} \geq \mu_0$, as \textit{endogenous fragility}. Figure \ref{fig:limit} shows the downstream average disruption probability $\bar{\mu}$ as a function of basal average disruption probability $\mu_0$, for cases in which basal correlation $\rho_0$ is low or high. In both cases for large possible initial levels of basal average disruption probability $\mu_0$ the supply chain is endogenously resilient, as $\bar{\mu} < \mu_0$. But, as in the uncorrelated cases studied above, there is a threshold of average basal disruption probability $\mu_0 > \mu^c$ for which the firm is endogenously fragile and $\bar{\mu} \geq \mu_0$. The threshold effect is discontinuous. At $\mu_0 \equiv \mu^c$ an arbitrarily small increase in $\mu_0$ can lead to discontinuously large downstream failure probabilities $\bar{\mu}$.

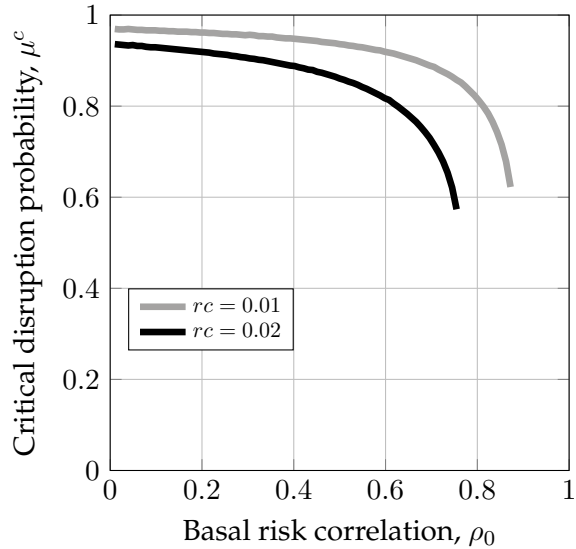
\begin{figure}[H]
  \centering
  % Recommended preamble:
\begin{tikzpicture}
\begin{axis}[width={0.5\linewidth}, height={0.5\linewidth}, ylabel={Critical disruption probability, $\mu^c$}, xlabel={Basal risk correlation, $\rho_0$}, grid={both}, ymin={0}, ymax={1}, xmin={0}, xmax={1}, legend style={at = {(0.4, 0.4)}, nodes={scale=0.75}}]
    \addplot[line width={2.5}, color={rgb,1:red,0.642;green,0.6405;blue,0.6372}]
        coordinates {
            (0.01,0.96942)
            (0.0198,0.96844)
            (0.0296,0.96844)
            (0.0394,0.96942)
            (0.0492,0.96844)
            (0.059,0.96746)
            (0.0688,0.96746)
            (0.0786,0.96648)
            (0.0884,0.96648)
            (0.0982,0.96648)
            (0.108,0.9655)
            (0.1178,0.9655)
            (0.1276,0.96452)
            (0.1374,0.96452)
            (0.1472,0.96354)
            (0.157,0.96354)
            (0.1668,0.96354)
            (0.1766,0.96256)
            (0.1864,0.96256)
            (0.1962,0.96158)
            (0.206,0.96158)
            (0.2158,0.9606)
            (0.2256,0.9606)
            (0.2354,0.95962)
            (0.2452,0.95864)
            (0.255,0.95864)
            (0.2648,0.95766)
            (0.2746,0.95766)
            (0.2844,0.95668)
            (0.2942,0.9557)
            (0.304,0.95668)
            (0.3138,0.9557)
            (0.3236,0.95374)
            (0.3334,0.95374)
            (0.3432,0.95276)
            (0.353,0.95276)
            (0.3628,0.9508)
            (0.3726,0.94982)
            (0.3824,0.94884)
            (0.3922,0.94884)
            (0.402,0.94786)
            (0.4118,0.94688)
            (0.4216,0.9459)
            (0.4314,0.94492)
            (0.4412,0.94394)
            (0.451,0.94296)
            (0.4608,0.941)
            (0.4706,0.94002)
            (0.4804,0.93904)
            (0.4902,0.93806)
            (0.5,0.9361)
            (0.5098,0.93512)
            (0.5196,0.93316)
            (0.5294,0.93218)
            (0.5392,0.93022)
            (0.549,0.92924)
            (0.5588,0.92728)
            (0.5686,0.92532)
            (0.5784,0.92336)
            (0.5882,0.92238)
            (0.598,0.91944)
            (0.6078,0.91748)
            (0.6176,0.91552)
            (0.6274,0.91258)
            (0.6372,0.90964)
            (0.647,0.9067)
            (0.6568,0.90376)
            (0.6666,0.90082)
            (0.6764,0.8969)
            (0.6862,0.89298)
            (0.696,0.88906)
            (0.7058,0.88612)
            (0.7156,0.88024)
            (0.7254,0.87534)
            (0.7352,0.87044)
            (0.745,0.86456)
            (0.7548,0.8577)
            (0.7646,0.85182)
            (0.7744,0.843)
            (0.7842,0.8332)
            (0.794,0.8234)
            (0.8038,0.81164)
            (0.8136,0.7989)
            (0.8234,0.7842)
            (0.8332,0.76558)
            (0.843,0.74304)
            (0.8528,0.7156)
            (0.8626,0.67836)
            (0.8724,0.6225)

        }
        ;
    \addlegendentry {$rc = 0.01 $}
    \addplot[line width={2.5}, color={rgb,1:red,0.0;green,0.0;blue,0.0}]
        coordinates {
            (0.01,0.9361)
            (0.0198,0.93512)
            (0.0296,0.93414)
            (0.0394,0.93316)
            (0.0492,0.93414)
            (0.059,0.93218)
            (0.0688,0.93218)
            (0.0786,0.93022)
            (0.0884,0.92924)
            (0.0982,0.92924)
            (0.108,0.92826)
            (0.1178,0.92728)
            (0.1276,0.9263)
            (0.1374,0.92532)
            (0.1472,0.92434)
            (0.157,0.92336)
            (0.1668,0.92238)
            (0.1766,0.9214)
            (0.1864,0.92042)
            (0.1962,0.91944)
            (0.206,0.91846)
            (0.2158,0.91748)
            (0.2256,0.91552)
            (0.2354,0.91454)
            (0.2452,0.91356)
            (0.255,0.91258)
            (0.2648,0.91062)
            (0.2746,0.90964)
            (0.2844,0.90768)
            (0.2942,0.9067)
            (0.304,0.90474)
            (0.3138,0.90376)
            (0.3236,0.9018)
            (0.3334,0.90082)
            (0.3432,0.89886)
            (0.353,0.8969)
            (0.3628,0.89494)
            (0.3726,0.89298)
            (0.3824,0.89102)
            (0.3922,0.88906)
            (0.402,0.88808)
            (0.4118,0.88514)
            (0.4216,0.88318)
            (0.4314,0.88024)
            (0.4412,0.87926)
            (0.451,0.87534)
            (0.4608,0.87338)
            (0.4706,0.87044)
            (0.4804,0.8675)
            (0.4902,0.86456)
            (0.5,0.86064)
            (0.5098,0.8577)
            (0.5196,0.85378)
            (0.5294,0.85084)
            (0.5392,0.84692)
            (0.549,0.84202)
            (0.5588,0.8381)
            (0.5686,0.8332)
            (0.5784,0.8283)
            (0.5882,0.8234)
            (0.598,0.81752)
            (0.6078,0.8136)
            (0.6176,0.80576)
            (0.6274,0.7989)
            (0.6372,0.79106)
            (0.647,0.78322)
            (0.6568,0.7744)
            (0.6666,0.76558)
            (0.6764,0.7548)
            (0.6862,0.74304)
            (0.696,0.7303)
            (0.7058,0.71462)
            (0.7156,0.69796)
            (0.7254,0.67836)
            (0.7352,0.65386)
            (0.745,0.62152)
            (0.7548,0.5735)

        }
        ;
    \addlegendentry {$rc = 0.02 $}
\end{axis}
\end{tikzpicture}
  \caption{Critical level of basal average disruption probability $\mu^c$ as a function of the basal correlation.}
  \label{fig:crit}
\end{figure}

Crucially, this threshold is decreasing in the basal level of correlation $\rho_0$, as illustrated in Figure \ref{fig:crit}. This implies that that even an arbitrary increase in basal correlation leads to discontinuously increases in the downstream average disruption probability. This results highlights an additional channel to that studied by \cite{elliott_supply_2022} by which supply chains can be endogenously fragile: even if the expected failure probability $\mu_0$ of basal producers remains unchanged, an increase in the correlation of their risk $\rho_0$, possibly due to climate change or offshoring, can endogenously induce large fragilities.

% Social Planner

\section{Social Planner Problem} \label{section:social-planner}

To establish a benchmark to which one can compare the competitive equilibrium analysed above, in this section I solve the model from the perspective of a social planner. The social planner attempts to, on the one hand, minimise the number of firms expected to fail, and, on the other, minimise the number of costly sourcing relations. To develop a useful benchmark, I define a social planner problem that can be meaningfully compared to the decentralised firm's problem by making the following two assumptions.

\begin{assumption} \label{assumption:social-planner:full-observation}
  The social planner knows the distribution of failure in the basal layer $P_0 \sim \Beta(\mu_0, \rho_0)$ and makes decision before $P_0$ is realised.
\end{assumption}

\begin{assumption} \label{assumption:social-planner:no-overlap}
  As in the firm problem, I consider the limit in which the number of firms producing each good goes to infinity, that is $n \to \infty$. This allows the social planner to recursively, from the last layer $K$ upwards, assign suppliers $\S_{k, i}$ such that there are sufficiently many firms so that no two firms share suppliers $\S_{k, i} \cap \S_{k, j} = \emptyset$.
\end{assumption}

\begin{wrapfigure}{r}{4cm}
  \centering
  \resizebox{\linewidth}{!}{\begin{tikzpicture}
	\begin{pgfonlayer}{nodelayer}
		\node [style=Dark blue firm] (01) at (2, 2) {};
		\node [style=Dark blue firm] (02) at (-1, 2) {};
		\node [style=Dark blue firm] (03) at (0, 2) {};
		\node [style=Dark blue firm] (04) at (1, 2) {};
		\node [style=Almost dark blue firm] (11) at (-1, 1) {};
		\node [style=Almost dark blue firm] (12) at (0, 1) {};
		\node [style=Almost dark blue firm] (21) at (-1, 0) {};
		\node [style=Almost dark blue firm] (22) at (0, 0) {};

		\node [draw, dotted, ultra thick, fit=(03), inner sep=4pt] (s03) {};
		\node [draw, ultra thick, fit=(01), inner sep=4pt] (s01) {};
	\end{pgfonlayer}
	\begin{pgfonlayer}{edgelayer}
		\draw [style=Supplies] (02) to (11);
		\draw [style=Supplies] (03) to (11);
		\draw [style=Supplies] (03) to (12);
		\draw [style=Supplies] (04) to (12);

		\draw [style=Supplies] (12) to (22);
		\draw [style=Supplies] (11) to (21);
	\end{pgfonlayer}
\end{tikzpicture}}
  \caption{}
  \label{fig:disentangle}
\end{wrapfigure}
To understand the intuition behind Assumption \ref*{assumption:social-planner:no-overlap}, consider the possible supplier overlap illustrate in Figure \ref{fig:disentangle}: if a supplier has multiple downstream clients (dashed box), the social planner can always rewire a link towards a supplier without downstream clients (solid box). By doing so, the social planner can ``diversify away'' all the correlation that arises due to the network structure. Hence, the only source of risk in the model is represented by the shutdowns experienced by firms in the basal layer, which happen with non-idiosyncratic probabilities $P_0$ (Assumption \ref{assumption:social-planner:full-observation}). Combining Assumptions \ref{assumption:social-planner:full-observation} and \ref{assumption:social-planner:no-overlap}, the social planner problem is then to maximise average expected payoffs

\begin{equation} \label{eq:social-planner:welfare}
  W(\{ \S_{k, i} \}) \coloneqq \frac{1}{K} \sum^{K}_{k = 0} \lim_{n \to \infty} \frac{1}{m_k(n)} \sum^{m_k(n)}_{i = 1} \Big(1 - \P\big( \mathcal{S}_{k, i} \subset \mathcal{D}_{k - 1} \big) \Big) \ \pi  - \frac{c}{2} \abs{\S_{k, i} }^2,
\end{equation} by choosing a sourcing strategy $S_{k, i} \subseteq \{1, 2, \ldots\}$ for each firm in each layer such that $S_{k, i} \cap S_{k, j}$ is empty for all $i, j$. The social planner problem can be further simplified by noticing that, given that all firms in layer $k$ are identical, if establishing an additional path from a firm in layer $k$ to a basal firm has positive marginal benefits, then it has positive marginal benefits for all firms in layer $k$ which share the same number of paths to basal firms. Hence, as in the decentralised firms' problem, the social planner can choose the optimal number of sources in each layer, let the firms source at random, and finally disentangle any overlapping paths. Using this, the social planner problem can be formulated recursively, by letting $V_k$ be the maximal average welfare from layer $k$ to the last layer $K$. This can be recursively defined as \begin{equation}
  V_k(P_{k - 1}) = \max_{s_k} \left\{ \left(1 - \mathbb{E}\big[ P_{k - 1}^{s_k} \big] \right) \pi -  \frac{c}{2} s_k^2 + \mathbb{E} \big[V_{k + 1}(P_k) \big] \right\}
\end{equation} where the state $P_{k - 1} \sim \BP(\mu_0, \rho_0, s_1 s_2 \ldots s_{k - 1})$ is the fraction of disrupted firms, which evolves as \begin{equation}
  P_k = P_{k - 1}^{s_k}.
\end{equation} The average welfare in layer $K + 1$ is given by $V_{K + 1}(P_K) = 0$, since firms in the last layer are never sources to other firms, and an initial state condition $P_0 \sim \Beta(\mu_0, \rho_0)$. This problem can be solved using standard backward induction techniques (see Appendix \ref{appendix:backward-induction}). The optimum average social welfare (\ref{eq:social-planner:welfare}) can then be written as \begin{equation}
  V_1(P_0) = \max_{s_1, s_2, \ldots s_{K - 1}} W(\{s_1, s_2, \ldots s_{K - 1}\}).
\end{equation}

\begin{figure}[H]
  \centering
  \input{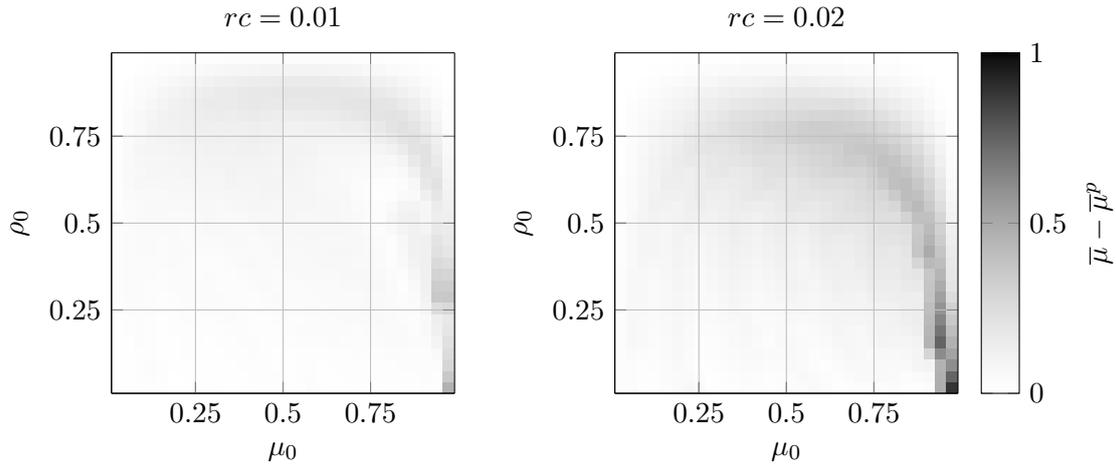}
  \caption{Change in downstream expected failure probability between the firms' $\overline{\mu}$ and the social planner $\overline{\mu}^p$ equilibrium, given different initial conditions $\mu_0$ and $\rho_0$ in a low (left) and a high (right) relative pairing costs regime.}
  \label{fig:limit-map:difference}
\end{figure}

Letting $\{s^p_k\}^K_{k = 1}$ be the socially optimal sourcing strategies sequence and $\{\mu^p_k\}^K_{k = 1}$ be the expected disruption in each layer given by such sourcing strategies, we can compute the change in downstream risk compared to the decentralised case. Figure \ref{fig:limit-map:difference} shows this difference $\overline{\mu} - \overline{\mu}^p$ for the same two cost regimes. If pairing costs are low, the social planner achieves marginally lower risk levels of downstream risk for most initial conditions. If initial basal correlation $\rho_0$ is sufficiently large and the average basal disruption probability $\mu_0$ is sufficiently low, the firms overdiversify compared to the socially optimum $\overline{\mu} < \overline{\mu}^p$. If relative pairing costs are high, the social planner is able to diversify risk around the critical threshold $\mu_c$, such that the decentralised equilibrium induces inefficiently high levels of average downstream disruption probability, that is $\overline{\mu} > \overline{\mu}^p$. This result implies that the cascading failures that occur around the critical threshold are fully attributable to firms' endogenous underdiversification motives and are hence inefficient.

\begin{figure}[H]
  \centering
  \input{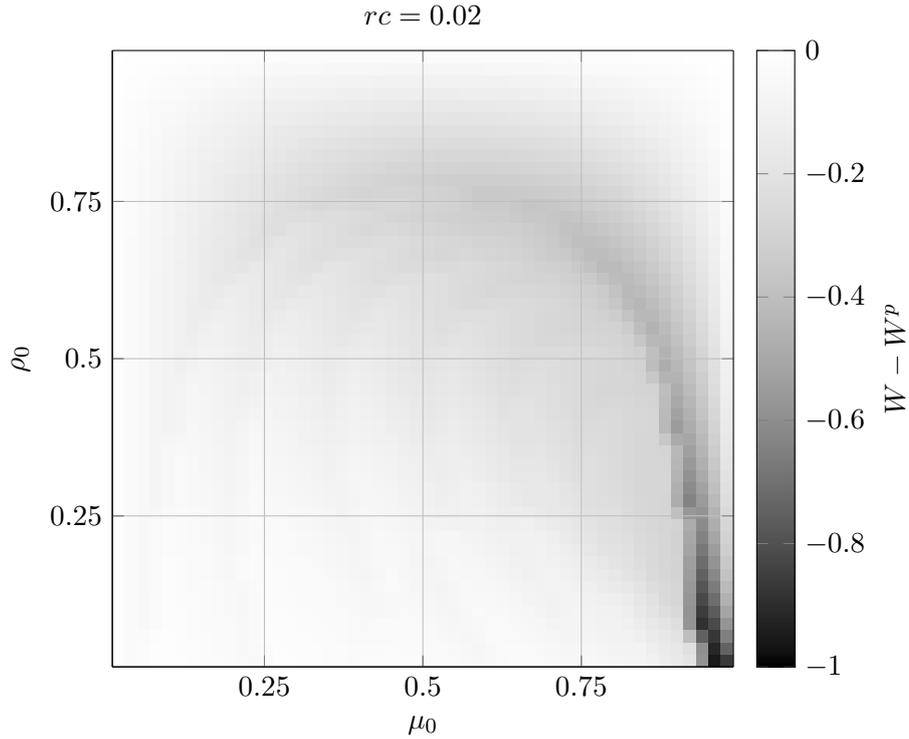}
  \caption{Welfare loss of the decentralised equilibrium $W - W^p$, given different initial conditions $\mu_0$ and $\rho_0$.}
  \label{fig:welfare-loss}
\end{figure}

The differences between the firms' sourcing strategies and the social optimum generate welfare losses in the production network. Letting $W$ be the average firm profit in the decentralised case and $W^p$ be the average profit achieved by the social planner, Figure \ref{fig:welfare-loss} illustrates the welfare loss due to the firms' diversification decisions $W - W^p$. The welfare loss is largest around the critical value $\mu^c$, where the production network is endogenously fragile. At these levels of risk, firms' upstream firms' diversification incentives are weak, which creates large downstream resilience externalities. Crucially, both an increase in basal risk $\mu_0$ and an increase in basal correlation $\rho_0$ can generate discontinuous welfare losses.

% Perfect information

\section{The Role of Opacity} \label{section:opacity}

So far I assumed that firms cannot observe the realisation of the supply chain and the basal disruption probabilities $P_0$ when making sourcing decisions. To understand how this assumption affects optimal decisions and fragility within the supply chain, I now analyse the supply chain under perfect information. The following assumption clarifies what is meant by perfect information in the context of the model.

\begin{assumption}
    In a regime of perfect information each firm firm $i$ in level $k$ is able to perfectly estimate the disruption probability of each potential supplier and the full correlation structure of the disruption events.
\end{assumption}

Under this perfect information regime, the firm can assign correct probabilities to its own disruption risk \begin{equation*}
    \mathbb{P} \Big( \S_{k, i} \subset \D_{k - 1} \Big) \text{ for all possible sourcing strategies } \S_{k, i}.
\end{equation*} The firm can hence rank suppliers by the marginal reduction in risk they provide and source from the ``safest'' $s_k$ desired suppliers. As all firms downstream are ex-ante identical, the marginal benefits of diversification experienced by firm $(i, j)$ are the same as those of all other firms in layer $k$, which implies that, in equilibrium, all firms in layer $k$ will employ the same sourcing strategy, given that they are ex-ante identical. This outcome is beneficial for any single firm, but detrimental for the stability of the production network. The following to propositions formalise this.

\begin{proposition}
    Compared to the opaque scenario, for the same number of sources, firm $(k, i)$ is (weakly) less likely to be disrupted.
\end{proposition}

\begin{proof}
    Given the same number of sources, the firm with perfect information minimises its disruption risk with fewer constraints than in the opaque scenario.
\end{proof}

\begin{proposition}
    Under perfect information, the supply chain is maximally fragile: either all firms fail or none do.
\end{proposition}

\begin{proof}
    The equilibrium outcome under perfect information implies that a disruption in layer $k$ affects all firm simultaneously as they all share the same sources, namely \begin{equation*}
        X_{k, i} = X_{k, j} \text{ for all } i \text{ and } j.
    \end{equation*} This, in turn, removes any diversification incentives downstream: a firm producing $k + 1$ cannot diversify its risk by multisourcing, hence it will single source.
\end{proof}

Opacity has a dual role: on the one hand, it prevents firms from optimally choosing the best sourcing strategy; on the other hand, it mitigates endogenous fragility by forcing firm to diversify.

% Conclusion

\section{Conclusion}

Risk diversification is a crucial determinant of firms’ sourcing strategies. In this paper, I show that firms endogenously underdiversify risk when they have incomplete information about upstream sourcing relations. This endogenous underdiversification generates fragile production networks, in which, arbitrarily small increases in correlation among disruptions between basal producers can generate discontinuously large disruptions downstream. I do so by deriving an analytical solution to a simple production game in which firms’ sole objective is to minimise the risk of failing to source input goods. 

Despite its simple structure, the game identifies an important externality firms impose on the production network when making sourcing decisions: upstream multisourcing introduces correlation in firms’ risk, which reduces incentives to multisource downstream. This externality exacerbates the risk of fragile production networks. Furthermore, I show that a social planner can design production networks that mitigate fully this externality. A consequence of this result is that, in principle, it is possible to design a transfer mechanism that allows downstream firms to compensate upstream firms to internalise the diversification externalities. By analysing the welfare loss in competitive equilibrium, I show that there is a critical region of basal conditions where arbitrary small increases in suppliers' correlation, even if the expected number of disrupted firms remains constant, can generate catastrophic downstream disruptions by altering downstream firm diversification incentives. This result illustrates how an increase in correlation among basal producers, for example due to widespread offshoring to the same country, can endogenously generate fragile production networks which have a large tail risk of disruption. Surprisingly, opacity plays a role in mitigating this effect, suggesting that if firms were to acquire information about the production network, despite the individual firm being better off, this could generate further endogenous fragility.

\iffalse
As mentioned before, the approach presented here is just one of the many points of view that can be taken when studying endogenous production network and imperfect information. Simple production games can be helpful in isolating mechanisms, but it is often equally, if not more, valuable to embed such mechanisms into more comprehensive and complex models and study how they interact with each other. In this spirit, it would be of great interest to develop a general equilibrium model with endogenous production network with growth and risk diversification motives. In addition, the social planner solution presented here serve as a natural steppingstone for the study of insurance or taxation schemes to mitigate production network externalities. How can we make funds flow upstream in the supply chain to incentives or disincentives diversification? Can such a transfer mechanism be setup without knowledge of the production network structure? The result presented in this paper suggest that this is possible. 
\fi

\newpage
\bibliography{supply-chain-reallocation}

\newpage
\appendix

\section{Notation and Distributions} \label{appendix:distribution-notation}

This appendix introduces standard notation and definitions that will be used throughout the following appendices. 

For $x \in \mathbb{R}$ and $n \in \mathbb{Z}$, I  denote the falling factorial as \begin{equation} \label{eq:falling-factorial}
  x^{\underline{n}} \coloneqq \underbrace{x(x - 1)(x - 2)\ldots(x - (n + 1))}_{n \text{ terms}}.
\end{equation} For non-integer exponents $s \in \mathbb{R}$, (\ref*{eq:falling-factorial}) can be extended as \begin{equation} \label{eq:falling-factorial:extension}
  x^{\overline{n}} \coloneqq \frac{\Gamma(x + n)}{\Gamma(x)},
\end{equation} where $\Gamma$ is the gamma function.

Two properties of the falling factorial that are used below but not proven are the additive property of the exponent \begin{equation} \label{eq:recursive-rising-factorial}
  x^{\overline{n + m}} = x^{\overline{n}} (x + m)^{\overline{m}},
\end{equation} and that it is strictly increasing in its base \begin{equation}
  \frac{\partial x^{\overline{n}}}{\partial x} > 0.
\end{equation}

\section{Omitted Proofs}

This appendix contains the proofs omitted from the paper.

\subsection[Proof of exchangeability]{Proof of Proposition \ref{proposition:exchangeability}} \label{appendix:proof-exchangeability}

Proving Proposition \ref{proposition:exchangeability}, requires the following Lemma. \begin{lemma} \label{lemma:set-size-matters}
  If the disruption events among upstream firms are exchangeable, then the probability that a downstream firm is disrupted depends only on the number of suppliers it picks.
\end{lemma}

\begin{proof}
  Consider the (possibly infinite) sequence of disruption events among upstream firms $X_{k, 1}, X_{k, 2}, X_{k, 3} \ldots $. We assume the sequence to be exchangeable, that is, \begin{equation}
    X_{k, 1}, X_{k, 2}, X_{k, 3}\ldots  \stackrel{d}{=} X_{k, \sigma(1)}, X_{k, \sigma(2)}, X_{k, \sigma(3)} \ldots ,
  \end{equation} for an arbitrary permutation of its indices $\sigma$. Fix two arbitrary finite subsets of firms $\mathcal{A} = \{X_{k, a_1}, X_{k, a_2} \ldots X_{k, a_n}\}$ and $\mathcal{B}= \{X_{k, b_1}, X_{k, b_2} \ldots X_{k, b_n}\}$ of size $n$. Here $a_i, b_i \in [m]$ are the indices of the original sequence corresponding to the $i$-th index of the subset. Let $\sigma$ be a permutation that takes elements of $\mathcal{A}$ to $\mathcal{B}$, namely, \begin{equation}
    \sigma(\mathcal{A}) = \mathcal{B} \text{ and } \sigma(\mathcal{A}^c) =  \mathcal{B}^c.
  \end{equation}

  Then the probability distribution over $\mathcal{A}$ is \begin{equation}
    \begin{split}
      \P(\mathcal{A}) &= \P(\mathcal{A} \text{ and } \mathcal{A}^c \text{ taking any value}) \\
      &= \P(\sigma(\mathcal{A}) \text{ and } \sigma(\mathcal{A}^c) \text{ taking any value}) \text{ then by exchangeability} \\
      &= \P(\mathcal{B} \text{ and } \mathcal{B}^c \text{ taking any value}) = \P(\mathcal{B}).
    \end{split}
  \end{equation}
\end{proof}

Now we can prove Proposition \ref{proposition:exchangeability}

\begin{proof}
  It can be proven by induction.

  The base case $k = 0$ follows from Assumption \ref{assumption:basal-exchangebility}, as the disruption probabilities are $\nu$-distributed and $\nu$ is symmetric.
  
  Assume that for some layer $k - 1$ the disruption events $X_{k - 1, 1}, X_{k - 1, 2} \ldots, X_{k - 1, m}$ are exchangeable. By Lemma \ref{lemma:set-size-matters} we know that the downstream expected profit $\Pi_{k, i}(\S)$ depends only the number of suppliers $\abs{\S}$. By symmetry and Assumption \ref{assumption:mixed-strategy}, all firms in layer $k$ are then selecting a random subset of supplier from layer $k - 1$ with equal probability, which in turn determines their disruption risk $X_{k, i}$. This construction is independent of the downstream firm index $i$, hence \begin{equation*}
    X_{k, 1}, X_{k, 2} \ldots, X_{k, m}
  \end{equation*} are exchangeable.
\end{proof}

\subsection{Proof of Proposition \ref{proposition:PtoP}} \label{appendix:PtoP}

\begin{proof}
  A firm producing good $k + 1$ sources from $s_{k + 1}$ suppliers, hence, its disruption event is given by \begin{equation}
    X_{i, k + 1} = X_{j_1, k} X_{j_2, k} \ldots X_{j_{s_{k + 1}}, k},
  \end{equation} where $\{j_1, j_2 \ldots j_{s_{k + 1}}\}$ is an arbitrary subset of suppliers and $X_{j, k}$ are exchangeable Bernoulli trials with a $P_k$ success probability, where \begin{equation*}
    P_k \sim \BP.
  \end{equation*} Conditional on the underline distribution $P_k$ of disruption probabilities, the trials $X_{j, k}$ are independent and identically distributed. Hence, we have \begin{equation}
    \begin{split}
      P_{k + 1} = \E[X_{i, k + 1}] &= \E\left[\prod^{s_k}_{l = 1} X_{j_l, k} \right] \\
      &= \E\left[\E\left[\prod^{s_k}_{l = 1} X_{j_l, k} \big\vert P_k = p_{j_l} \right]\right] \text{ by conditional independence, } \\
      &= \E\left[ \E\left[ X_{j_l, k} \big\vert P_k = p_{j_l} \right]^{s_k} \right] \text{ by independence of the draws } p_{j, l} \\
      &= \E\left[ \E\left[ X_{j_l, k} \big\vert P_k = p_{j_l} \right] \right]^{s_k} = P_k^{s_k}.
    \end{split}
  \end{equation}
\end{proof}

\subsection{Mapping of risk across layers} \label{appendix:derivations}

This section derives the risk reduction factor $\eta$.

\begin{lemma} \label{lemma:final-step}
  If $P_{k - 1} \sim \BP(m, \alpha, \beta, S)$ for some integer $S$, then \begin{equation}
    P_{k} \sim \BP(m, \alpha, \beta, S s_k)
  \end{equation} where $s_k$ is the choice of suppliers in layer $k$.
\end{lemma}

\begin{proof}
  Follows from the definition of $\BP$.
\end{proof}

\begin{proposition} \label{proposition:risk}
  The expected probability of disruption faced by a firm is given by \begin{equation}
    \E \Big[ P_k \Big] = \frac{B\left(\mu_0 \; \frac{1 - \rho_0}{\rho_0} + S_{k - 1} s_k, (1 - \mu_0) \; \frac{1 - \rho_0}{\rho_0} \right)}{B\left(\mu_0 \; \frac{1 - \rho_0}{\rho_0}, (1 - \mu_0) \; \frac{1 - \rho_0}{\rho_0} \right)}.
  \end{equation}
\end{proposition}

\begin{proof}
  It follows from rewriting the moment generating function of the beta distribution as \begin{equation}
    M(t) = \sum^\infty_{n = 0} \frac{t^n}{n!} \frac{B(\alpha + n, \beta)}{B(\alpha, \beta)}.
  \end{equation} and noticing that $P_k \sim \BP(\alpha, \beta, S_{k - 1} s_k)$.
\end{proof}

To simplify notation, let $r_0 = \frac{1 - \rho_0}{\rho_0}$ and \begin{equation} \label{eq:eta-definition}
  \eta(s, S) = \frac{\left(\mu_0 r_0 + S\right)^{\overline{S(s - 1)}}}{\left(r_0 + S\right)^{\overline{S(s - 1)}}} 
\end{equation} which satisfies the recursion \begin{equation} \label{eq:eta-recursion}
  \eta(s + 1, S) = \eta(s, S) \frac{(\mu_0 r_0 + Ss)^{\overline{S}}}{(r_0 + Ss)^{\overline{S}}}.
\end{equation} Another property of (\ref{eq:eta-definition}) which will be use later is \begin{equation} \label{eq:eta-derivative}
  \frac{\partial \eta}{\partial s} = \eta(s, S) S \; \Big(\psi(\mu_0 r_0 + S s) - \psi(r_0 + S s) \Big).
\end{equation}

\begin{corollary} \label{corollary:decrease-of-eta}
  From equation (\ref{eq:eta-derivative}) and the fact that $\psi$ is increasing over positive values, it follows that $\eta$ is decreasing in $s$.
\end{corollary}

Using Proposition (\ref{proposition:risk}), the coefficient $\eta$ allows us to write the propagation of risk recursively \begin{equation}
  \mu_{k + 1} = \eta(s_{k + 1}, S_k) \; \mu_{k}.
\end{equation}

\subsection[Limit case]{Limit case $\rho_0 \to 0$} \label{appendix:limit_case}

For the following proof I only consider non-trivial values of upstream risk $\mu < \mu^{0}$. If $\mu \geq \mu^0$, no firm has suppliers and the supply chain is by definition stable.

\begin{lemma}
  A fixed point of the law of motion $g(\Bar{\mu}) = \Bar{\mu}$, is attained iff $\tilde{g}(\Bar{\mu}) \geq \Bar{\mu}$.
\end{lemma}

\begin{proof}
  By definition $\tilde{g}(\Bar{\mu}) = \Bar{\mu}^{\tilde{s}(\Bar{\mu})}$ and $0 \leq \Bar{\mu} \leq 1$. Hence $\tilde{g}(\Bar{\mu}) \geq \Bar{\mu} \iff \tilde{s}(\Bar{\mu}) \in (0, 1]$. By definition $s(\Bar{\mu}) = \ceil{\tilde{s}(\Bar{\mu})} = 1$, which implies that $g(\Bar{\mu}) = \Bar{\mu}$.
\end{proof}

Now we can prove Corollary \ref{corollary:basin:no-corr}.

\begin{proof}
  We seek $\mu$, such that $\tilde{g}(\mu) \geq \mu$, which then implies that $g(\mu) = \overline{\mu}$. This will be the case if $\tilde{s}(\mu) \in (0, 1]$. This is the case if $\Delta\Pi(1) \leq 0$ and $\Delta\Pi(0) > 0$, which yields the desired inequality.
\end{proof}

\subsection[General Case]{General Case, $\rho_0 > 0$} \label{appendix:general-case-proofs}

This appendix proves the existence of an optimal sourcing in the case $\rho > 0$.

\begin{proof}
  It is sufficient to show that $\Delta\Pi$ is strictly decreasing in $s$ when $\rho_0 > 0$. It is convenient to rewrite $\eta$ (\ref{eq:eta-definition}) as \begin{equation}
    \eta(s) = \frac{\Gamma(r_0 + S)}{\Gamma(\mu_0\; r_0 + S)} \frac{\Gamma(\mu_0\; r_0 + S s)}{\Gamma(r_0 + S s)}.
  \end{equation} Then \begin{equation}
    \Delta\Pi(s) = \big(\eta(s) - \eta(s + 1)\big) \pi \mu - c \left(s + \frac{1}{2}\right), 
  \end{equation} hence \begin{equation}
    \begin{split}
      \Delta\Pi'(s) = - c -\mu \pi S \Bigg( &\eta(s + 1) \Big( \psi(\mu_0 r_0 + S(s + 1)) - \psi(r_0 + S(s + 1)) \Big) - \\
      &\eta(s) \Big( \psi(\mu_0 r_0 + S s) - \psi(r_0 + S s) \Big) \Bigg).
    \end{split}
  \end{equation}

  Then $\Delta\Pi'(s) < 0$, since $\psi$ is increasing. Finally notice that $\Delta\Pi(-1/2) =( \eta(-1/2) - \eta(1/2)) \pi \mu < 0$ and $\lim_{s \to \infty} \Delta\Pi(s) = \infty$.
\end{proof}

After proving that a solution exists, I will now prove that it is concave in the level of correlation $\rho_0$ (Proposition \ref{proposition:concavity-of-incentives}).

\begin{proof}
  Notice that another way of writing $\Delta \Pi$ is letting $P \sim \BP(\mu, \rho, s)$ and writing \begin{equation}
    \Delta\Pi(s) = \E\Big[P^s - P^{s - 1}\Big] \pi - c \left(s + \frac{1}{2}\right).
  \end{equation} The optimal incentive $s$, is a root of $\Delta\Pi$, hence \begin{equation}
    s = \frac{\E\Big[ P^s - P^{s - 1} \Big]}{c} \pi \mu - \frac{1}{2} = \frac{\pi \mu}{c} \Big( \E[P^s] - \E[P^{s - 1}] \Big) - \frac{1}{2}.
  \end{equation}

  % \textbf{TODO: This should be obvious but I cannot see it now.}
\end{proof}

\section{Solution of the Social Planner Problem} \label{appendix:backward-induction}

First, notice that the terminal condition $V_K$ is linear in $P_{K - 1}$, hence \begin{equation}
  \E\big[V_K(P_{K - 1})\big] = V_K\big(\E[P_{K - 1}]\big).
\end{equation} In turn, this implies that $V_k$ is linear for all $k$. Hence we can rewrite the value to be a function of the state space $S$, \begin{equation}
  V_{k}(S_{k - 1}) = \max_{s} \Bigg\{ \Big(1 - \E\big[\BP(\mu_0, \rho_0, S_{k - 1} \; s)\big] \Big) \pi - \frac{c}{2} s^2 + V_{k}(S_{k - 1} \; s_k) \Bigg\}.
\end{equation}

We can find $V_k$ numerically. Let $\Omega = [m] \times [m^K]$ for some $m \in \mathbb{N}$ and \begin{equation}
  l(s, S) \coloneqq \Big(1 - \E\big[\BP(\mu_0, \rho_0, S_{k - 1} \; s)\big] \Big) \pi - \frac{c}{2} s^2.
\end{equation}

Then we can recursively compute backward \begin{equation}
  \begin{split}
    V_K(S) &= \max_s l(\Omega), \\
    V_{k - 1}(S) &= \max_s l(\Omega) + V_K(S \; s), \\
    \vdots \\
    V_{1}(S) &= \max_s l(\Omega) + V_2(S \; s).
  \end{split}
\end{equation}

\end{document}